\documentclass{PoS}

\def\schpt{S\raise0.4ex\hbox{$\chi$}PT}
\def\cO{{\cal O}}
\def\ltwid{{\,\raise.3ex\hbox{$<$\kern-.75em\lower1ex\hbox{$\sim$}}\,}}
\def\MeV{{\rm Me\!V}}
\def\chpt{\raise0.4ex\hbox{$\chi$}PT}
\def\msbar{{\overline{\rm MS}}}
\def\GeV{{\rm Ge\!V}}

\title{Update on the physics of light pseudoscalar mesons}

\ShortTitle{Update on the physics of light pseudoscalar mesons}

\author{C. Bernard\\
        Washington University; Saint Louis, Missouri, USA}

\author{C. DeTar\\
        University of Utah; Salt Lake City, Utah, USA}

\author{Steven Gottlieb and L. Levkova\footnote{Current address:
University of Utah; Salt Lake City, Utah, USA}\\
        Indiana University; Bloomington, Indiana, USA}

\author{U.M.\ Heller\\
        American Physical Society; Ridge, New York, USA}

\author{J.E.\ Hetrick\\
        University of the Pacific; Stockton, California, USA}

\author{J. Osborn\\
        Boston University; Boston, Massachusetts, USA}

\author{D. Renner and D. Toussaint\\
        University of Arizona; Tucson, Arizona, USA}

\author{\speaker{R. Sugar}\\
        University of California; Santa Barbara, California, USA}

\abstract{
We present an update of the MILC investigation of the properties of light
pseudoscalar mesons using three flavors of improved staggered quarks.
Results are presented for the $\pi$ and $K$ leptonic decay constants,
the CKM matrix element $V_{us}$, the up, down and strange quark masses,
and the coefficients of the $\cO(p^4)$
chiral lagrangian.  We have new data for lattice 
spacing $a \approx 0.15$~fm with several values of the light quark mass down to
one-tenth the strange quark mass,
higher statistics for $a\approx 0.09$~fm with the 
light quark mass equal to one-tenth the strange quark 
mass, and initial results for our smallest lattice spacing, $a \approx 0.06$~fm
with light quark mass two-fifths of the strange quark mass.}

\FullConference{XXIVth International Symposium on Lattice Field Theory\\
                July 23-28, 2006\\
                Tucson, Arizona, USA}

\begin{document}

\section{Introduction}

We present an update of our long term investigation of properties of $\pi$
and $K$ mesons~\cite{FPI04,FPILAT04,FPILAT05} with three flavors of
improved staggered quarks~\cite{ASQTAD}. Study of the $\pi-K$ system is 
interesting because it enables: 1) a sensitive check of algorithms
and methods, including the use of the fourth root of the determinant
for dynamical staggered quarks, by comparing lattice results for $f_\pi$ to the 
well-determined experimental value; 2) a precise extraction of the CKM 
matrix element $V_{us}$ from $f_K$ or $f_K/f_\pi$, competitive with the 
world-average from alternative methods; 3) a determination of the light 
quark masses and their ratios with high precision; 4) a calculation of 
several of the
physical coefficients of the ${\cal O}(p^4)$ chiral Lagrangian, the
Gasser-Leutwyler $L_i$; 5) a test of the applicability of staggered chiral 
perturbation theory (\schpt)~\cite{LEE-SHARPE,SCHPT-CACB,SCHPT-OTHER}
for describing lattice data; and 6) a determination of the extra, unphysical 
parameters that enter \schpt.  This allows, for example,
use of heavy-light \schpt~\cite{HL-SCHPT} for computations of decay
constants and  form factors in $D$ and $B$ systems~\cite{FERMI-HL} 
without introducing large additional uncertainties.

\section{Gauge Configurations and Chiral Fits}

The MILC Collaboration has generated an extensive set of gauge
configurations with three flavors of improved staggered quarks:
$m'_u\!=\!m'_d\!\equiv\! \hat m'$, and $m'_s$. (Primes indicate simulation 
values; corresponding masses without primes are the physical values).
In this work we report results for ensembles with lattices spacings
$a\approx 0.15$, 0.12, 0.09 and 0.06~fm, which are referred to as
coarser, coarse, fine and super-fine lattices, respectively. Coarser,
coarse and fine lattices are available with a range of light quark masses,
the lowest pion mass being $m_\pi\!\approx\! 240\; \MeV$.  
A run on the super-fine lattices with $m_\pi\!\approx\! 430\; \MeV$ 
is half-finished, and lighter-mass runs are being started.  The simulation 
strange quark masses $m'_s$ are in the range 
$0.70\, m_s \ltwid m'_s \ltwid 1.2\, m_s$.  The physical volumes of the 
lattices 
range from $\approx\!(2.4\;{\rm fm})^3$ to $\approx\!(3.4\;{\rm fm})^3$, 
with the large volumes being used for ensembles with the lightest quark masses. 
The lattice spacing is kept fixed within each ensemble (coarser, coarse,
fine and super-fine) as the light quark mass is varied, using the length
$r_1$~\cite{SOMMER,MILC-POTENTIAL} from the static quark potential
to set the scale.  The absolute scale is set from the $\Upsilon$ $2S$-$1S$  
splitting determined by the HPQCD Collaboration~\cite{DAVIES,PRL} on most of 
our lattices. From their results we find $r_1 = 0.318(7)$~fm.

For Goldstone masses and decay constants, we have
extensive partially quenched data, typically all combinations
of 8 or 9 valence quark masses between $0.1\,m'_s$ and $m'_s$. Goldstone
quantities have the smallest statistical errors, so we concentrate on
them. We fit decay constants and masses together including all correlations, 
and we fit different 
lattices spacings together. The confidence level of the joint fit is 0.99.
Statistical errors are very small: $0.1\% $ to $ 0.8\%$ for squared 
masses, and $0.1\% $ to $ 1.4\%$ for decay constants. 

In order to obtain good fits to \schpt\ forms, we need to place upper limits 
on quark masses. We consider two data sets:

\noindent
{\bf Subset I:} In this subset the valence Goldstone pion masses are restricted
to be $\ltwid 350\; \MeV$, and the other pion masses to be $\ltwid 550\; \MeV$ 
for
the coarse ensemble, $\ltwid 460\; \MeV$ for the fine ensemble, and 
$\ltwid 400\; \MeV$ for the super-fine ensemble. The largest sea quark masses
and the coarser ensemble are excluded, leaving a total of 122 data points.
We expect errors of 
NLO \schpt\ to be of order 0.8\%, implying that NNLO terms are needed.
NNLO \schpt\ logs are unknown;  however, for higher masses, where NNLO 
terms are important, such logs should be smoothly varying and well 
approximated by NNLO analytic terms. The NNLO fit has 20 unconstrained
parameters. An additional 6 tightly constrained parameters allow for 
variation of physical LO and NLO parameters  with the lattice spacing 
($\sim\! \alpha_Sa^2\Lambda_{\rm QCD}^2\approx 2\%$), giving a total of 
26 parameters. This subset is used to determine the 
$L_i$, and the systematic errors on other quantities.  

\noindent
{\bf Subset II:} In this subset the valence Goldstone pion masses are
restricted to be $\ltwid 570\; \MeV$, and the other pion masses to be
$\ltwid 780\; \MeV$ for the coarser ensemble, $\ltwid 725\; \MeV$ for the coarse
ensemble, $\ltwid 630\; \MeV$ for the fine ensemble, and $\ltwid 590\; \MeV$
for the super-fine ensemble. All sea quark masses and the coarser ensemble are
included, leaving 978 data points. Even NNLO fits break down for this subset.
We therefore add 18 NNNLO analytic terms and 10 tightly constrained variations 
of NNLO parameters with $a$. We fix (within errors) LO and NLO terms from the 
fit to subset I. We are then left with 28 unconstrained parameters and  26 
tightly constrained ones.  This data set is not chiral, but it is
used to interpolate around the physical value of $m_s$. The central values of 
decay constants and quark masses are determined from it.

We emphasize that the necessity for using high order \schpt\ fits with large
numbers of parameters arises from the very small statistical errors
in our data. If we did not care about the confidence level, we could perform
a NLO \schpt\ fit with only 12 parameters. It changes $f_\pi$ by 4\%, 
$f_K$ by 1\%, and  $m_s$ by 3\% on Subset I; however, it has 
$\chi^2/d.o.f.=9.5$ for 110 $d.o.f.$. Similarly, the NLO \schpt\ fit with
taste-violating parameters given as input and the lattice-spacing dependence 
of physical parameters set to zero has 6 parameters. It changes $f_\pi$ by 2\%,
$f_K$ by 1\% and $m_s$ by 0.5\% for Subset I; but it has
$\chi^2/d.o.f.=40.5$ for 110 $d.o.f.$. Furthermore, such fits can change
the low energy constants by 3 or 4~$\sigma$. It is important to note that
if the physics is not correct, one will not be able to fit the data even with
$\sim\!40$ parameters. This was illustrated by a series of fits in 
Ref.~\cite{FPI04}. There, comparable fits to the continuum \chpt\ form
had 36 parameters, but $\chi^2/d.o.f.= 8.8$ for 204 $d.o.f.$, for
a confidence level of $10^{-250}$. Fits with all 
chiral logs 
and finite volume corrections omitted from the fit function (analytic function 
only) were poor, providing good evidence for chiral logs. Finally, 
separate linear fits for $m_\pi^2$ or $f_\pi$ as a function of the average
valence quark mass were tried.  For $m_\pi^2$ there were 6 parameters and  
$\chi^2/d.o.f.\approx 20$ for 234 $d.o.f.$, while for $f_\pi$ there were
10 parameters, and $\chi^2/d.o.f.  \approx 25$ for 230 $d.o.f.$.

\section{Taste Symmetry and Chiral Logs}

\begin{figure}[t]
\begin{center}
\parbox[t]{0.45\linewidth}{
\centerline{\includegraphics[width=\linewidth]{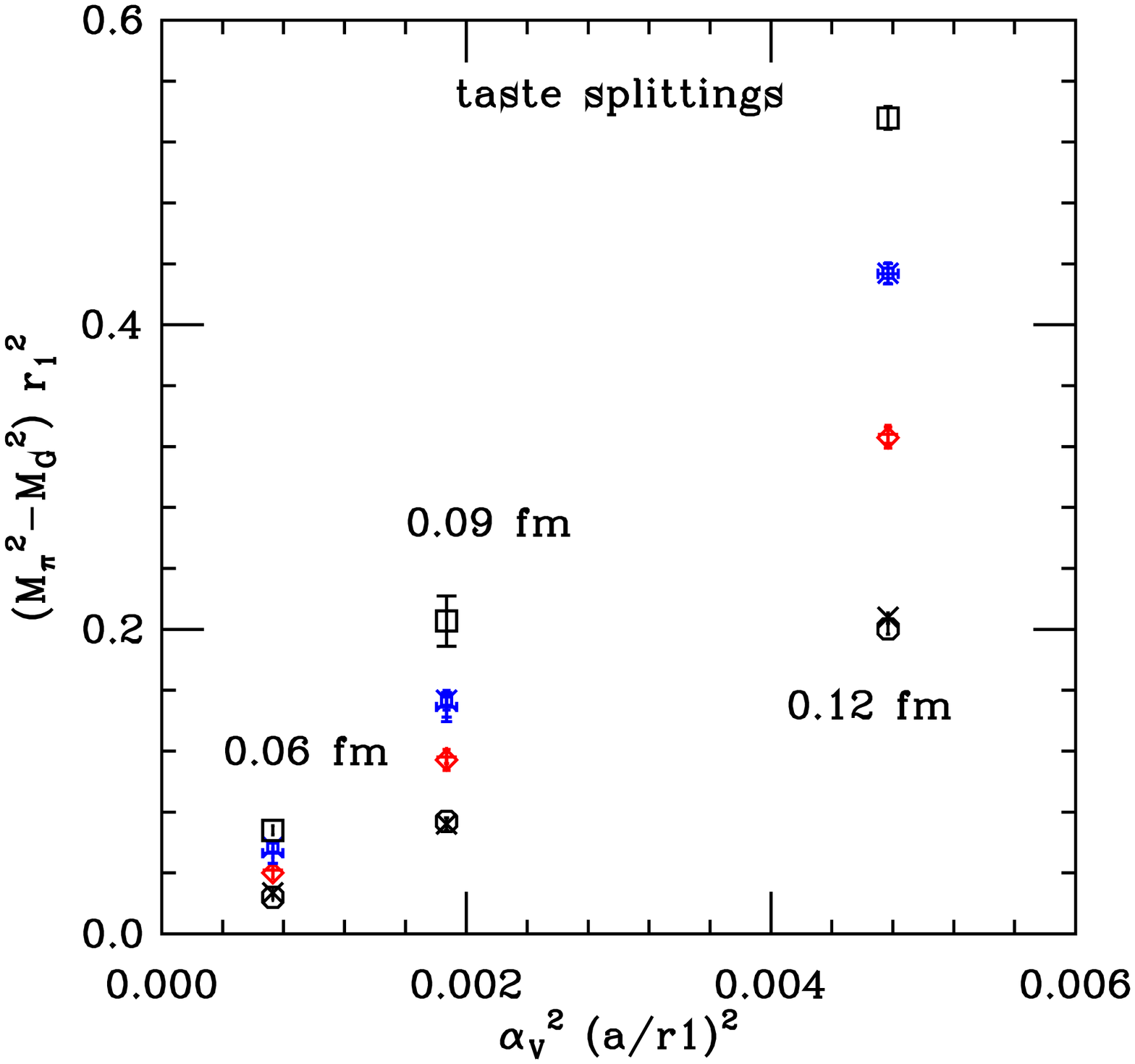}}
\caption{Taste splitting among the pions for lattice spacings 0.12, 0.09
and 0.06 fm with $\hat m'=0.4\, m_s'$.}
\label{TASTE}
}
\hspace{0.05\linewidth}
\parbox[t]{0.45\linewidth}{
\centerline{\includegraphics[width=0.94\linewidth]{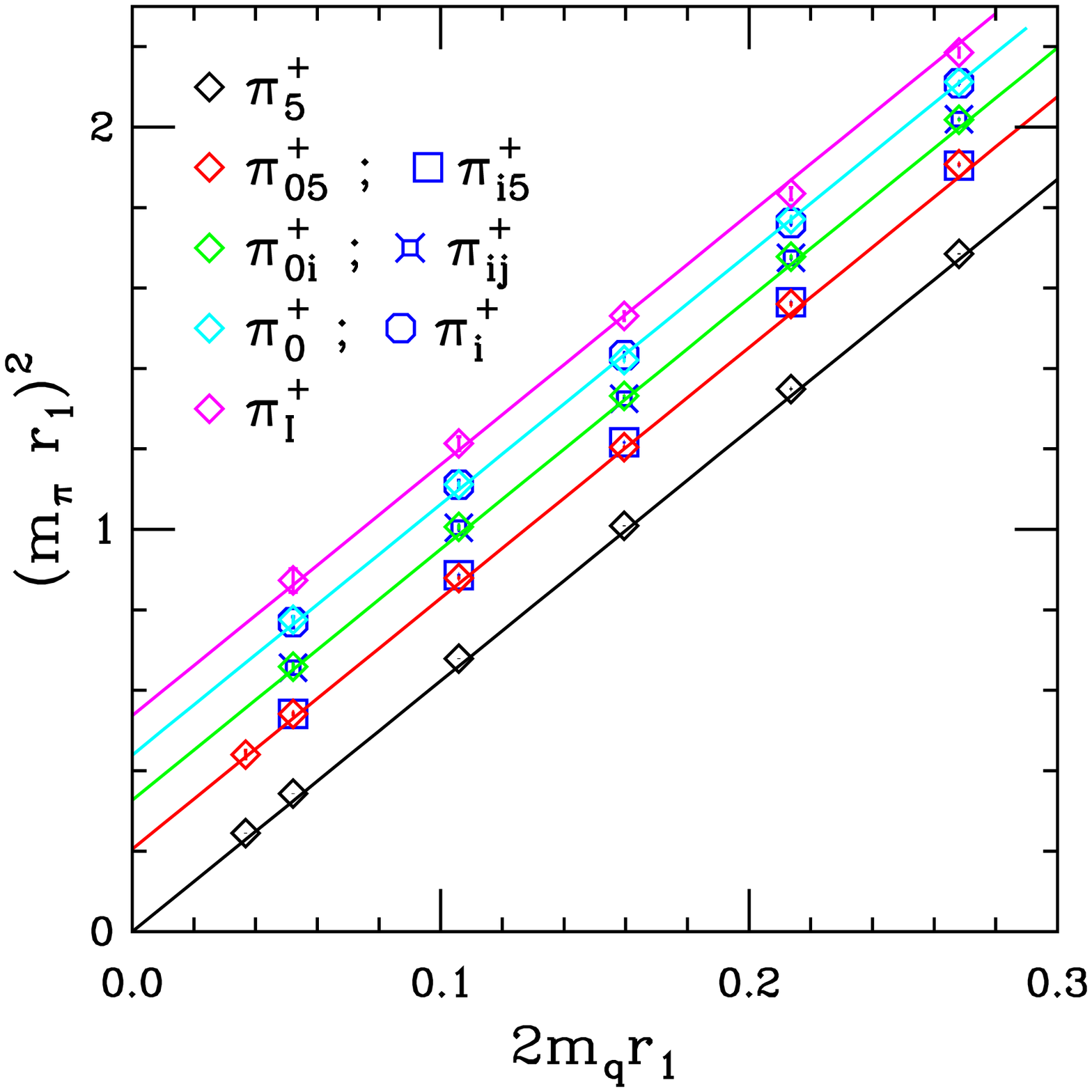}}
\caption{Squared pion masses as a function of the light quark mass on
the coarse lattices.}
\label{FULL}
}
\end{center}
\end{figure}

Violations of taste symmetry for staggered quarks are most apparent in the
pion sector. Figure~\ref{TASTE} 
shows the splitting between the square of Goldstone pion mass and the
square of the masses of pions with other tastes for the coarse,
fine and super-fine ensembles with $\hat m'=0.4\, m'_s$. The splittings
are plotted as a function of $a^2 \alpha_s^2$, the variable in which they
are expected to be linear at small lattice spacings. Figure~\ref{FULL} shows
the square of the pion masses as a function of the light quark mass. 
The degeneracies predicted by Lee and Sharpe in Ref.~\cite{LEE-SHARPE} are
clearly visible. These two figures provide strong evidence that taste
symmetry violations are well understood from \schpt, and that they
decrease as expected with decreasing lattice spacing.

The data is accurate enough and at small enough quark masses so that the
effects of chiral logs are evident. This is illustrated in Fig.~\ref{LOGS}
where the square of the pion mass divided by the sum of the valence quark
masses $m_x+m_y$ is plotted as a function of 
$m_x+m_y$ in units of $r_1$. The mass renormalization 
constant, relative to that of the  fine lattices, has been included so that 
data from all lattice spacings can be presented on the same plot. Lines 
through the data points come from the \schpt\ fit to the entire data set for 
decay constants and masses discussed above.
The upward slope in these lines at small valence quark masses is caused
by partially quenched chiral logs, which are of the form 
$(m_{sea,G}^2 +\Delta_i a^2 \alpha_S^2) \log(m_{val,G}^2 +
\Delta_i a^2 \alpha_S^2)$, where $m_{val,G}$ and $m_{sea,G}$  are the 
valence and sea Goldstone pion masses, respectively, and 
$\Delta_i$ parametrizes the splitting of the taste-$i$ pion.
The effect is especially pronounced
in the super-fine data because, not only are the taste-violating mass
splittings rather small, but the light sea quark mass is rather 
large, $\hat m'=0.4\, m_s'$.
The red line is the fit function in ``full continuum QCD'' (valence and sea
quark masses set equal, and extrapolation of the parameters to the continuum
limit).  It is much smoother because it does not have partially quenched
chiral logs.

\section{Preliminary Results}

Our previously published work~\cite{FPI04,FPILAT04,FPILAT05} was based on
analysis of the coarse and fine ensembles. Here we present preliminary results
obtained by including the coarser and super-fine ensembles, and additional
data from the $\hat m'=0.1\, m'_s$ run in the fine ensemble.

\begin{figure}[t]
\begin{center}
\parbox[t]{0.45\linewidth}{
\centerline{\includegraphics[width=0.97\linewidth]{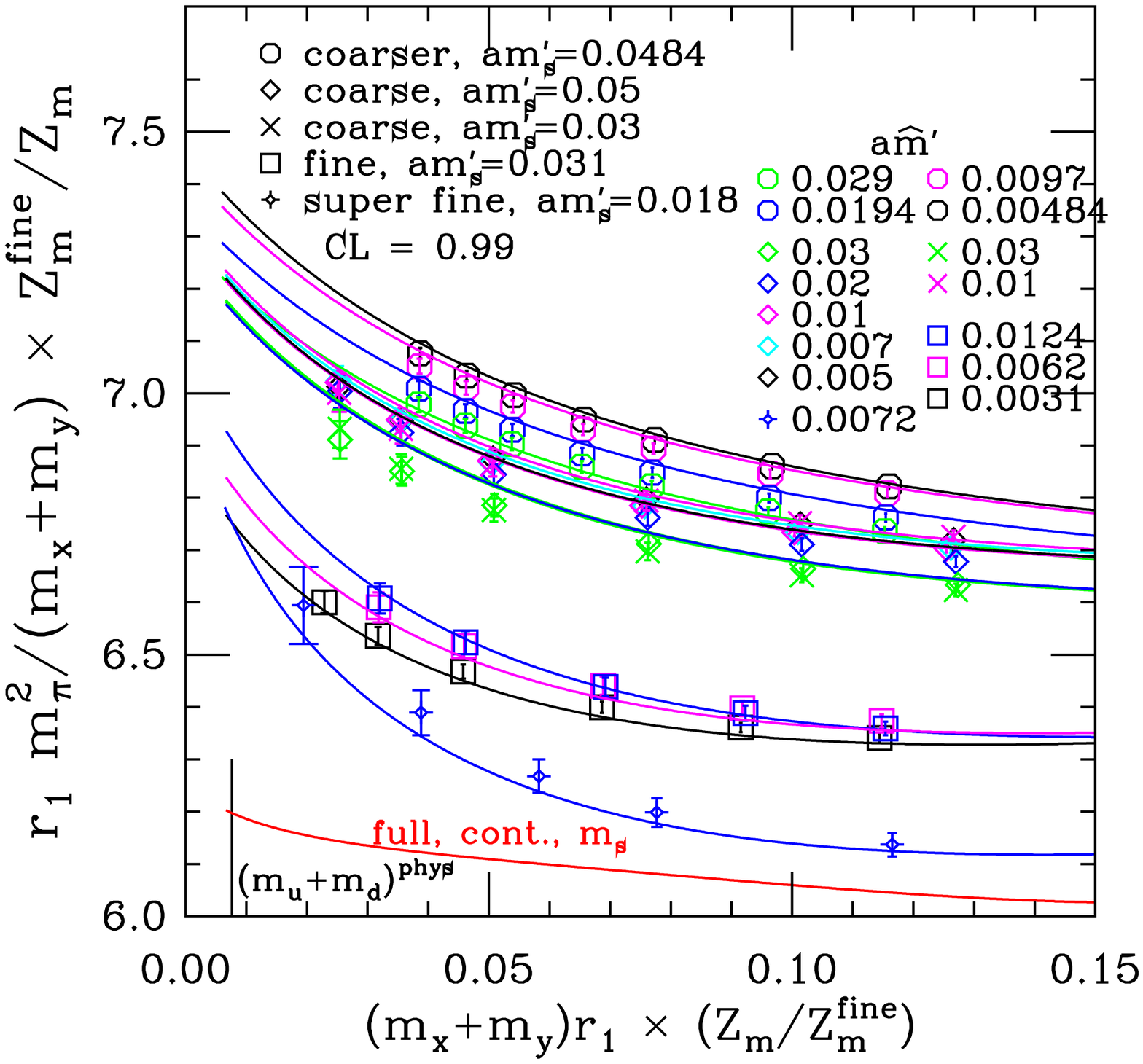}}
\caption{The square of the pion mass divided by the sum of the valence quark 
masses as a function of the sum of the valence
quark masses in units of $r_1$.}
\label{LOGS}
}
\hspace{0.05\linewidth}
\parbox[t]{0.45\linewidth}{
\centerline{\includegraphics[width=\linewidth]{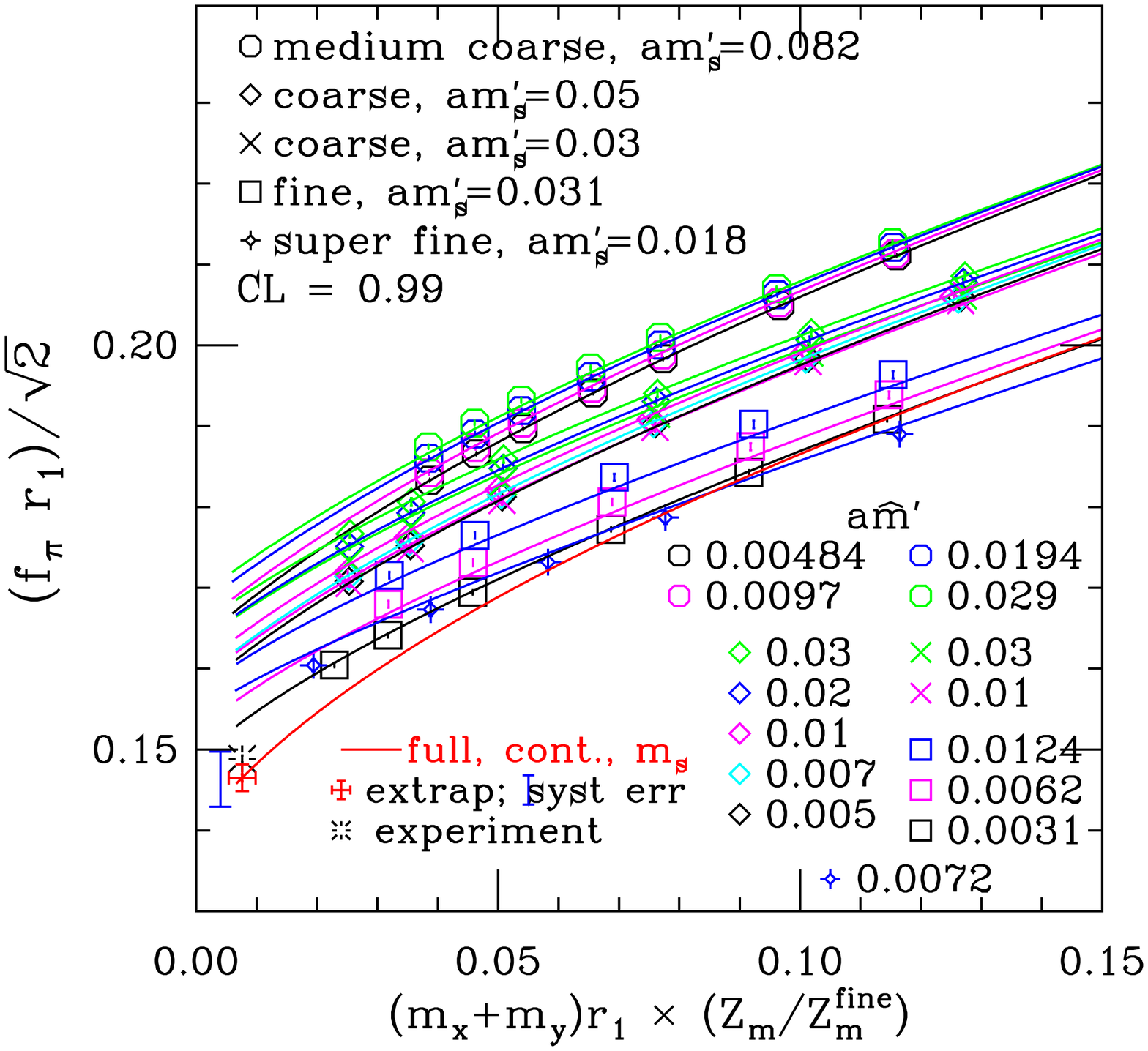}}
\caption{The pion decay constant as a function of the sum of the valence
quark masses in units of $r_1$.}
\label{FPI}
}
\end{center}
\end{figure}

Figure~\ref{FPI} shows fit for the leptonic decay constant of the pion
as a function of the
sum of the valence quark masses, in units of the scale $r_1$. Lines
through the data points come from the same \schpt\ fit as in Fig.~\ref{LOGS}.
The red lines represent the fit function in ``full continuum QCD''
with the strange sea quark mass fixed to its physical value.
The red plus shows the extrapolated value of $f_\pi$ to the physical point, 
in agreement with experiment (black burst) to within systematic errors 
(blue bar).

Preliminary numerical results for $f_\pi$ and $f_K$ obtained from our most
recent data are

\begin{eqnarray}\label{eq:f_results}
f_\pi  =  128.6 \pm 0.4\pm 3.0 \; \MeV && [{129.5 \pm 0.9\pm 3.5 \; \MeV}] \\
f_K  =    155.3 \pm 0.4\pm 3.1 \; \MeV &&  [{156.6 \pm 1.0\pm 3.6 \; \MeV}] \\
f_K/f_\pi   = 1.208(2)({}^{+\phantom{1}7}_{-14}) &&   [{ 1.210(4)(13)}].
\end{eqnarray}

Here the numbers on the left are the new values, and those on the
right in square brackets are
from Refs.~\cite{FPI04,FPILAT04,FPILAT05}. In each case the
first error is statistical, and the second systematic. The result for $f_\pi$
obtained from the experimental rate for $\pi\!\to\!\mu\nu$
coupled with the value of $V_{ud}$ from super allowed nuclear beta decay is
$f_\pi=130.7 \pm0.1\pm 0.4\; \MeV$. The agreement of our result with the
experimental one provides important evidence that we do understand and
can control our errors.  Marciano has pointed out~\cite{MARCIANO} that
lattice results for $f_K/f_\pi$ can be combined with the experimentally
determined rate for $K\!\to\!\mu\nu$ and $V_{ud}$ to calculate $V_{us}$.
We find

\begin{eqnarray}
|V_{us}|=0.2223({}^{+26}_{-14}) &&    [0.2219(26)].
\end{eqnarray}

Again, our latest result is on the left with the previous one in square
brackets on the right. 
The Particle Data Group (2006) gives $V_{us}\!=\!0.2257(21)$~\cite{PDG}
from the $K\!\to\!\pi\mu\nu$ experimental rate and non-lattice theory.
Lattice errors continue to dominate experimental ones in our determination 
of $V_{us}$, and they will be reduced as additional super-fine lattices 
become available.

The up, down and strange quark masses can be determined from the
masses of the $\pi$ and $K$ mesons using the \schpt\ fits. To do so
we must distinguish between experimental masses, QCD masses in which
electromagnetism has been turned off, and those in which both electromagnetism
and isospin violations are turned off. The last of these, which we denote
by $m_{\hat\pi}$ and $m_{\hat K}$, are used to determine the physical values
of the bare quark masses $\hat m$ and $m_s$ from the \schpt\ fits. They are 
related to the QCD masses by

\begin{eqnarray}
m^2_{\hat\pi} & \approx & (m^{QCD}_{\pi^0})^2\\
m^2_{\hat K} & \approx & [(m^{QCD}_{K^0})^2 + (m^{QCD}_{K^+})^2]/2.
\end{eqnarray}

To relate the QCD masses to the experimental ones requires some continuum
input~\cite{CONT}. Details can be found in Ref.~\cite{FPI04}. Once
$\hat m$ and $m_s$ have been determined, $m_u$ can be obtained from
$m^{QCD}_{K^+}$. Preliminary results from our current data set are:

\begin{eqnarray}
m_s^\msbar = 90(0)(5)(4)(0)\;\MeV &&  [76(0)(3)(7)(0)\;\MeV]\\
\hat m^\msbar = 3.3(0)(2)(2)(0)\;\MeV &&   [2.8(0)(1)(3)(0)\; \MeV] \\
m_s/\hat m = 27.2(0)(4)(0)(0) &&  [27.4(1)(4)(0)(1)] \\
m_u^\msbar = 2.0(0)(1)(1)(1)\;\MeV &&  [1.7(0)(1)(2)(2)\;\MeV] \\
m_d^\msbar = 4.6(0)(2)(2)(1)\;\MeV  &&   [3.9(0)(1)(4)(2)\;\MeV] \\
m_u/m_d = 0.42(0)(1)(0)(4)  && [0.43(0)(1)(0)(8)]
\end{eqnarray}

Again, new results are on the left, and early ones in this case from
Refs.~\cite{FPI04,PRL,STRANGE-MASS} are in square brackets on the right.
Errors are from statistics, simulation systematics, perturbation theory and
electromagnetic effects. The $\msbar$ scale is $\mu\!=\!2\,\GeV$. 
The main difference between the new and old
results comes from the perturbation theory calculation of the renormalization
constants needed to match the lattice masses to the $\msbar$~masses. 
Here we use the new two-loop results of Mason, Trottier and Horgan~\cite{MTH},
whereas the old results were based on a one-loop calculation. A non-perturbative
mass renormalization calculation is in progress. Note that our result for
$m_u/m_d$ rules out a massless $m_u$ at the $10\, \sigma$ level.

Finally, we have determined a number of the low energy constants $L_i$
at the scale of $m_\eta$. Our preliminary results, in units of $10^{-3}$,
are:

\begin{eqnarray}\label{eq:Li_results}
2L_6 - L_4 = 0.5(1)(2) && [0.5(2)(4)] \\
2L_8 - L_5 = -0.1(1)(1) && [ -0.2(1)(2)] \\
L_4 = 0.1(2)(2) && [ 0.2(3)(3)] \\
L_5 =2.0(3)(2) && [1.9(3)(3)]. 
\end{eqnarray}

Again, the new values are on the left, and the older ones from Ref.~\cite{FPI04}
are in square brackets on the right.
These results are consistent with ``conventional'' ones summarized in
Ref.~\cite{CKN}: $L_5 = 2.2(5) $, $L_6 = 0.0(3)$, and $L_4 = 0.0(5)$.
Our value for $2L_8-L_5$ is far from the range $-3.4 \le 2L_8-L_5 \le -1.8 $
that would allow $m_u=0$, which is consistent with, but not independent
of, our direct determination of $m_u$. Our old results for the low energy
constants already play a significant role in setting constraints on the 
$\pi-\pi$ scattering lengths $a^2_0$ and $a^0_0$~\cite{CCL}. It will be 
interesting to see the impact of the new ones.

\noindent
{\bf Acknowledgments:} The work was supported in part by the Department of
Energy and the National Science Foundation. Computational resources were
provided by FNAL, IU, NCSA, NERSC, PSC, ORNL, and SDSC.

\end{document}